\documentclass{emulateapj}        

\def\etal{{\it et~al. }}

\begin{document}

\title{Performance Modeling of a Wide Field Ground Layer Adaptive Optics System}
\shorttitle{Performance Modeling of a Wide Field GLAO System}

\author{David R. Andersen\altaffilmark{1}, Jeff Stoesz\altaffilmark{1},
Simon Morris\altaffilmark{2}, Michael Lloyd-Hart\altaffilmark{3},
David Crampton\altaffilmark{1}, Tim Butterley\altaffilmark{2},
Brent Ellerbroek\altaffilmark{4}, Laurent Jolissaint\altaffilmark{1},
N. Mark Milton\altaffilmark{3}, Richard Myers\altaffilmark{2},
Kei Szeto\altaffilmark{1}, Andrei Tokovinin\altaffilmark{5},
Jean-Pierre V\'eran\altaffilmark{1}, Richard Wilson\altaffilmark{2}}

\affil{
\altaffilmark{1}NRC Herzberg Institute of Astrophysics, 5071 W Saanich Road,
Victoria, BC V9E 2E7, Canada, \\
\altaffilmark{2}Department of Physics, University of Durham, Rochester
Building, Science Laboratories, South Road, Durham DH1 3LE, UK,\\
\altaffilmark{3}Steward Observatory, University of Arizona, 933 North Cherry
Avenue, Tucson, AZ 85721, USA, \\
\altaffilmark{4}Thirty Meter Telescope Project, 1200 E. California Boulevard,
Mail Code 102-8, Pasadena, California, 91125, USA,\\
\altaffilmark{5}Cerro Tololo Inter-American Observatory, Casilla 603, La
Serena, Chile
}
\email{david.andersen@cnrc-nrc.gc.ca}

\begin{abstract}

Using five independent analytic and Monte Carlo simulation codes,
we have studied the performance of wide field ground layer
adaptive optics (GLAO), which
can use a single, relatively low order deformable mirror to correct
the wavefront errors from the lowest altitude turbulence. 
GLAO concentrates more light from a point source in a
smaller area on the science detector, 
but unlike traditional adaptive optics, images
do not become diffraction-limited. Rather the GLAO point spread function (PSF)
has the same functional form as a seeing-limited PSF,
and can be characterized by familiar performance metrics such as Full-Width
Half-Max (FWHM).
The FWHM of a GLAO PSF is reduced by
0.1$^{\prime\prime}$ or more for optical
and near-infrared wavelengths over different atmospheric conditions.
For the Cerro Pach\'on atmospheric model this correction
is even greater when the image quality is worst, which effectively eliminates
``bad-seeing'' nights; the best seeing-limited image quality, available only
20\% of the time, can
be achieved 60 to 80\% of the time with GLAO. This concentration of 
energy in the PSF will reduce required exposure times
and improve the efficiency of an
observatory up to 30 to 40\%. These performance gains are relatively
insensitive to a number of trades including the exact field of view of 
a wide field GLAO system, the conjugate altitude and actuator density
of the deformable mirror, and the number and configuration of the guide
stars. 

\end{abstract}


\section{Introduction}

The idea that the turbulence of the atmosphere can be corrected
by adaptive optics (AO) is not new (Babcock 1953), but there have
always been limitations to the approach. One of the more serious
is that classical or single guide star AO systems produce
only a small corrected field of view (FOV); isoplanatic errors cause the
image quality to quickly degrade from the 
center of the corrected field; typically, the spatial resolution falls below
the diffraction limit in the near-infrared only
30 arcseconds from the guide star.
This small and non-uniform corrected FOV severely limits the sky coverage
of traditional AO systems, and even limits sky coverage for those that employ
Laser Guide Stars (LGSs; first proposed by Foy \& Labeyrie 1985). 
Another limitation of traditional AO systems
is that the performance decreases from the near-infrared to the visible;
to fully correct the turbulence in the optical would require
deformable mirrors (DMs) with many more actuators and a control
system operating at a much higher frequency (Dekany 2006). Finally, existing
AO systems perform well only when the image quality conditions are
good; if the seeing is poor, control bandwidths and 
DM actuator strokes and densities are insufficient to maintain
diffraction limited imaging.

In the near future,
Multi-Conjugate Adaptive Optics (MCAO) systems (Johnston \& Welsh 1991;
Ragazzoni, 1999; Beckers, 2000;
Ragazzoni, Farinato,
\& Marchetti 2000; Flicker, Rigaut \& Ellerbroek 
2000) promise to produce
a larger corrected FOV with improved sky coverage,
by employing multiple DMs and wavefront
sensors (WFSs), but will still be limited to observations in the
near-infrared on nights with average or better image quality.
While the availability of MCAO systems will change the
use of observatories,
it is important to realize that today a large fraction
of ground based astronomical research still relies on seeing limited
observations at optical wavelengths. 
Ground layer adaptive optics (GLAO) was proposed to circumvent these
limitations of traditional AO systems by applying a 
limited AO correction to an even larger
FOV under any atmospheric conditions even at optical
wavelengths (Rigaut 2002).
A GLAO system does not attempt to
produce diffraction limited images, but rather to
improve the concentration of the point spread function (PSF)
by sensing and correcting only the lowest
turbulent layers of the atmospheres. Because the corrected layers are
so close to the ground, the correction is the same over the entire large
FOV. Uncorrected turbulent layers at higher altitudes 
degrade
the spatial resolution isoplanatically. The use of GLAO can therefore complement
MCAO surveys; the survey efficiency of a GLAO system, as we show, continues
to increase as the FOV increases and actually surpasses that of a MCAO
system. MCAO or classical AO can then be used for follow-up observations
of individual discoveries made from GLAO surveys.

Early simulations suggested that GLAO could produce images with a 
Full-Width Half Maximum (FWHM) less than 0.2$^{\prime\prime}$ in $J$-band.
Caution should be used when evaluating early GLAO modeling results,
however, because they are critically dependent on the input turbulence
profiles. As Tokovinin (2004) points out, studies of GLAO require
accurate knowledge of the atmosphere below $\sim$2 km in addition to knowing
the turbulence profile of the free atmosphere. Specifically,
turbulent layers at intermediate altitudes which Tokovinin termed the 
``gray zone,'' will only be partially corrected and will also introduce
residual anisoplanatism. Earlier work on GLAO focused on 
FOVs less than 3 arcminutes (Baranec, Lloyd-Hart, Codona, 
Milton 2003; Tokovinin 2004; LeLouarn \& Hubin 2004; Hubin \etal 2004; 
Jolissaint, V\'eran, \& Stoesz  2004; Stoesz \etal 2004). 

Here we examine the performance of a GLAO
system over FOVs greater than 5 arcminutes in size using 
new, higher resolution turbulence
profiles.
This work was carried out to study the potential of a GLAO system
at the {\it Gemini} Observatory, but the results are applicable to 10m-class
telescopes in general. The results presented here are based on a set of model
atmospheres that were derived at least in part from balloon measurements of
the turbulence over
Cerro Pach\'on which had a vertical resolution of 6m (Tokovinin \&
Travouillon 2006).
After describing these model atmospheres in section 2, we present the
analytic and Monte Carlo modeling tools used for this study in section 3.
We also reconcile the results of these codes,
which lends
greater confidence in the results.
Section 4 describes the GLAO PSF and relevant
performance metrics. We apply these metrics in analyzing a baseline wide field
GLAO system in section 5 and describe various trades on this baseline
in section 6. 
Finally section 7 provides a summary of our modeling results and a 
discussion of the promising future of GLAO.

\section{Model Atmospheres}

GLAO system performance depends crucially on the structure
of the atmospheric turbulence profile. In particular, the size
of the compensated field and the uniformity of the delivered
PSF over the field depend on the thickness of the boundary layer,
while the overall degree of image improvement depends sensitively 
on the ratio of aberrations in the boundary layer to those in the
free atmosphere. Historically these are not quantities that
have been studied in detail because, prior to the
emergence of GLAO as a potentially valuable observing tool,
they were not seen as important measures of a site's quality.

Fortunately for our study,
detailed measurements of the
structure of the atmospheric turbulence with a resolution of 6m in the boundary layer
(altitudes below 5000m)
recorded from 43 balloon flights exist as part of the 1998 {\it Gemini-South} 
seeing campaign at Cerro Pach\'on 
(Vernin \etal 2000).
The atmospheric 
turbulence profiles
used in this study have been derived from those balloon data and
MASS--DIMM data also from Cerro Pach\'on taken in 2003
(Tokovinin \& Travouillon 2006; see also Tokovinin, Baumont, \& Vasquez 2003). 
While more
site testing is required to confirm the presence of strong ground
layer turbulence at other sites,
the results of GLAO simulations using
the high resolution atmospheric data from Cerro Pach\'on should be 
generally applicable to other 
telescope sites.

In order to reduce the magnitude of the modeling task,
a total of nine atmospheric profiles have been constructed from the
balloon flight data that represent a broad range of typical
conditions. Three turbulence
profiles for the atmosphere below 2km in altitude were computed,
representing the averages of the 25\% best, 25\% worst, and central
50\% of the data sorted by $r_0$. Similarly the free atmosphere above 2 km was
represented by a single layer of turbulence, with $C_n^2$ and height
determined again from the best (25\%), worst (25\%), and 
typical (50\%) conditions for the upper atmosphere. Essentially no
degree of correlation between the strengths of the boundary
layer and the free atmosphere were found (Tokovinin 2003), so nine
profiles were constructed by matching the three boundary layer 
profiles with each of the three upper atmosphere layers. Table 1 gives
the altitudes and integrated turbulence values ($J =\int C_n^2 {\rm d}h$ in m$^{1/3}$)
for each of these profiles. The first column of this
table is the effective height of each layer above the site level,
as defined by the integral $\int h C_n^2 {\rm d}h / J$, 
and the remaining columns 
are the integrated turbulence values 
for each model profile. The lowest layer was assigned a height of zero in this
study.
The approximate probabilities of 
occurrence, given the percentiles that the profiles were drawn from, are
given in Table 2. Figure 1 shows good agreement between the cumulative
probability distribution from the model predictions to
measurements made from {\it Gemini-South}.
Hence the results of our simulations should be representative of the
conditions at Cerro Pach\'on during the four, one-week balloon missions that
took place in all four seasons of 1998. We refer to these nine atmospheric
models as two word (or two letter) designations for the Ground and Free
atmosphere profiles, respectively. Hence, Good-Typical refers to the model
atmosphere consisting of the ``Good'' (i.e. largest $r_0$) ground layer profile
and ``Typical'' (median $r_0$) free atmosphere profile. A caveat to the results
presented here is that while the profiles exhibit a wide range of $r_0$
values, almost all have similar isoplanatic angles, $\theta_0$; in all
cases we assumed an outer scale of $L_0 = 30$m (Table 3).

\begin{deluxetable}{llll}
\tabletypesize{\small}
\tablewidth{0pt}
\tablecaption{Integrated turbulence $J =\int C_n^2 {\rm d}h$ in m$^{1/3}$ 
for ``Good,'' ``Typical,'' and ``Bad''
Ground and Free (altitudes$>3$km are considered ``Free'') atmospheres.}
\tablehead{
\multicolumn{1}{l}{altitude} &
\multicolumn{1}{l}{Good} &
\multicolumn{1}{l}{Typical} &
\multicolumn{1}{l}{Bad} \\
\multicolumn{1}{l}{(m)} &
\multicolumn{1}{l}{$J$ (10$^{-14}$ m$^{1/3}$)} &
\multicolumn{1}{l}{$J$ (10$^{-14}$ m$^{1/3}$)} &
\multicolumn{1}{l}{$J$ (10$^{-14}$ m$^{1/3}$)} 
}
\startdata
0 &     9.26 &  7.04 &  13.8 \\
25 &    1.83 &  2.25 &  10.8 \\
50 &    0.574 & 1.35 &  15.3 \\
100 &   0.362 & 1.24 &  15.8 \\
200 &   0.614 & 1.99 &  10.3 \\
400 &   0.960 & 2.87 &  6.46 \\
800 &   1.18 &  3.02 &  7.29 \\
1600 &  0.913 & 1.75 &  6.77 \\
\hline
3600 & & & 32.0 \\
5500 & & 17.0 & \\
8400 &  9.00 & & 
\enddata
\end{deluxetable}

\begin{deluxetable}{llll}
\tabletypesize{\small}
\tablewidth{0pt}
\tablecaption{Probabilities used to weight atmospheric models}
\tablehead{
\multicolumn{1}{l}{Free} &
\multicolumn{3}{c}{Ground Layer} \\
\multicolumn{1}{l}{Atmosphere} &
\multicolumn{1}{c}{Good} &
\multicolumn{1}{c}{Typical} &
\multicolumn{1}{c}{Bad} 
}
\startdata
Good & 0.0625 & 0.125 & 0.0625 \\
Typical & 0.125 & 0.250 & 0.125 \\
Bad & 0.0625 & 0.125 & 0.0625 
\enddata
\end{deluxetable}

\begin{figure}
\vbox to 2.2in{\rule{0pt}{2.2in}}
\includegraphics{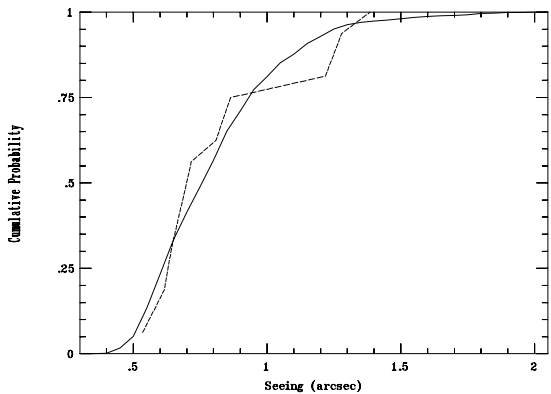}
\caption{Cumulative histogram of {\it Gemini-South} seeing measurements 
in the $R$-band, with
probabilities of the model atmospheric profiles over-plotted (dashed line).
The measurements were made from {\it Gemini-South} acquisition
camera data corrected to zenith (Data available at 
http://www.gemini.edu/metrics/seeing.html). There is good agreement between
these curves; the probabilities of the derived atmospheric profiles
are within present experimental uncertainty.}
\end{figure}

\begin{deluxetable*}{llllllllll}
\tabletypesize{\small}
\tablewidth{0pt}
\tablecaption{Values of $r_0$ in meters, and $\theta_0$ and seeing
in arcseconds
at 500nm for
all nine C$_n^2$ profiles used in this study.}
\tablehead{
\multicolumn{1}{l}{} &
\multicolumn{9}{c}{Ground Layer} \\
\multicolumn{1}{l}{Free} &
\multicolumn{3}{c}{Good} &
\multicolumn{3}{c}{Typical} &
\multicolumn{3}{c}{Bad} \\
\multicolumn{1}{l}{Atmosphere} &
\multicolumn{1}{l}{$r_0$} &
\multicolumn{1}{l}{$\theta_0$} &
\multicolumn{1}{l}{FWHM} &
\multicolumn{1}{l}{$r_0$} &
\multicolumn{1}{l}{$\theta_0$} &
\multicolumn{1}{l}{FWHM} &
\multicolumn{1}{l}{$r_0$} &
\multicolumn{1}{l}{$\theta_0$} &
\multicolumn{1}{l}{FWHM} \\
\multicolumn{1}{l}{} &
\multicolumn{1}{l}{m} &
\multicolumn{1}{l}{$^{\prime\prime}$} &
\multicolumn{1}{l}{$^{\prime\prime}$} &
\multicolumn{1}{l}{m} &
\multicolumn{1}{l}{$^{\prime\prime}$} &
\multicolumn{1}{l}{$^{\prime\prime}$} &
\multicolumn{1}{l}{m} &
\multicolumn{1}{l}{$^{\prime\prime}$} &
\multicolumn{1}{l}{$^{\prime\prime}$} 
}
\startdata
Good    & 0.189 & 2.61 & 0.535 & 0.164 & 2.59 & 0.616 & 0.083 & 2.52 & 1.218 \\
Typical & 0.157 & 2.72 & 0.644 & 0.141 & 2.70 & 0.717 & 0.079 & 2.62 & 1.279 \\
Bad     & 0.125 & 2.84 & 0.809 & 0.117 & 2.82 & 0.864 & 0.073 & 2.73 & 1.385
\enddata
\end{deluxetable*}

\section{Modeling Tools}

In assessing the expected performance of GLAO, five simulation codes written by
four groups have been used. These were thoroughly tested and compared to
each other to ensure a high degree of confidence in the results.
Three codes implement analytic calculations, while the remaining two codes are 
full wave propagation Monte Carlo simulations.
Analytic codes, which calculate an 
estimate of the long exposure AO-corrected PSF using the fact that
the optical transfer function is proportional to the
negative exponential of the aperture averaged structure
function of the residual phase disturbances in the telescope pupil,
were generally used to explore large
parameter spaces and study various performance trades. 
Monte Carlo models are much more computationally intensive, and
were primarily used to study physical effects not
incorporated into analytic models and to verify the analytic model
results for the baseline configurations.

\subsection{PAOLA Analytic Modeling Tool}

One of the analytic modeling codes, PAOLA (Performance of Adaptive
Optics for Large Apertures; Jolissaint, V\'eran 2002; 
Jolissaint, V\'eran, Conan 2006) was developed
at NRC-HIA and is now used by more than a dozen groups throughout the world.
It models the effect of the AO correction as a spatial frequency filtering
of the turbulent phase power spectrum, from which the AO long exposure PSF
in any direction is easily derived. One can identify five basic limitations on
any classical AO system, and these are taken into account in the PAOLA code:
\begin{description}
\item[Anisoplanatism:] in GLAO mode the DM commands are assumed to be
derived from an average of the multiple guide star WFS measurements. The difference
between this average command and the actual turbulent phase at a given point in the
field is called the {\it anisoplanatic error}, and is described in
(Stoesz \etal 2004) for the case of multiple natural guide star GLAO. It is important
to note that this error term is by far the most important for GLAO;
\item[Fitting error:] The number of WFS lenslets (and/or DM actuators) defines
the number of aberration modes that can be corrected by the system, and in
particular sets the highest spatial frequency that can be measured and corrected.
Uncorrected high spatial frequency aberrations are transmitted to the output
of the AO system, giving rise to what is called the {\it fitting error},
due to the limited ability of the system to adjust (fit) itself to the
incident phase. Fitting error is the next most important source of residual 
aberrations for GLAO;
\item[WFS spatial aliasing:] These same uncorrected high frequency aberrations
are seen by the WFS as low spatial frequency errors, and are aliased in the
low-frequency domain of the WFS. {\it Aliasing error} is the third 
important GLAO error source;
\item[WFS noise:] This error is due to the guide stars photon noise and WFS
detector read noise and dark current noise;
\item[System servo-lag:] To achieve sufficient signal-to-noise on
the phase measurement, the WFS has to integrate for a given exposure
time. Determining the exposure time involves a 
trade-off between getting enough guide star photons and
averaging out the high temporal phase fluctuations. Moreover, the
reading of the WFS, the phase reconstruction and the DMs surface
update takes some time (roughly one sampling period), creating a
time-lag between phase measurement and correction. The phase error
term associated to both time averaging and time-lag is called the
{\it servo-lag error};
\end{description}
The AO loop controller is modeled in PAOLA as a simple integrator. Such an
analytical approach is very computationally efficient and permits
PAOLA to model AO performance across large parameter spaces in a 
reasonable period of time. However, it can
only account for the fundamental limits of the AO correction,
so the performance estimates need to be refined with much more
computationally intensive Monte-Carlo simulation tools (see sections 3.4 and
3.5), taking into account non-linear and/or second order effects (correlated effects,
sensitivity to vibrations, cone effect, spot elongation, etc.), 
once a reduced set of suitable parameters
has been found. 

\subsection{CIBOLA}

CIBOLA (Covariance-Including Basic Option for Linear Analysis; 
Ellerbroek 2005) is a second analytical modeling tool that combines
and extends features of PAOLA and prior analytical models for tomographic
wavefront reconstruction and MCAO (Tokovinin 2001; Tokovinin 2002). This
code may be used to assess the correlated effect of five fundamental error
sources (DM fitting error; WFS spatial aliasing; WFS measurement noise;
finite servo bandwidth; and anisoplanatism) for AO systems incorporating 
one or more deformable mirrors and wavefront sensors. Narrow and wide-field
performance estimates may be obtained in terms of wavefront error power
spectra and point spread functions, computed using either conventional,
MCAO, or GLAO control algorithms.

The principal capabilities and limitations of CIBOLA are derived from the
use of spatial filtering approximations for all of the basic wavefront
propagation, sensing, reconstruction, and correction operators encountered
in classical linear systems modeling of adaptive optics. This approach
enables rapid analysis of AO systems which is sufficiently accurate for
many applications, but is also neglects aperture edge effects and is 
(rigorously) limited to the case of natural guide stars.

\subsection{Arizona analytic code}

The idl-based analytic GLAO simulation tool used at the University of Arizona
was originally developed by Tokovinin (2004).
A multi-layer residual structure
function for each beacon is computed from the von Karman power spectrum
at each turbulent layer, accounting for the geometry of the beacon
constellation.  The model assumes that a single natural guide star
is used for sensing global tilt.
The effects of temporal delay and wavefront sensor noise are neglected.

\subsection{Arizona Monte Carlo simulation code}

The Monte Carlo simulation tool written at the University of
Arizona, described by Lloyd-Hart \& Milton (2003), supports an
arbitrary number of LGS, natural guide stars (NGS), 
DMs, and atmospheric turbulence layers.
The model assumes the geometric optics approximation.  Atmospheric
turbulence and DM corrections are represented as vectors of coefficients
of the Zernike modes.  An analytic computation is used to obtain the
influence of atmospheric turbulence at each layer within the intersecting
cone for each LGS and NGS.  The net aberration for an object at infinity
is also computed analytically.

The reconstruction matrix is built from the product of the maximum {\it a
priori} (MAP) inverse
of the DM influence matrix and the atmospheric layer influence matrix.
Random turbulence Zernike coefficient vectors are generated from the
Cerro Pach\'on atmospheric models using Kolmogorov statistics for Zernike
order $1\le n \le 30$ ({\it i.e.}, the first 496 Zernike polynomials).  
DM corrections are the product of these
random turbulence Zernike coefficient vectors and the reconstruction
matrix.  Read noise with Gaussian statistics and Poisson photon noise
are simulated for a Hartmann-Shack WFS and added to the noise-free
wavefront corrections.

The performance of each candidate GLAO beacon configuration is evaluated
by calculating the RMS wavefront deviation at a range of different field
positions out to the full field radius.  The expected uncorrected RMS error
for Zernike modes of order $n > 30$ is added to account for high frequency
modes not included in the simulation.

\subsection{Durham Monte Carlo adaptive optics model}

The University of Durham  Monte Carlo AO model includes detailed
WFS noise propagation and produces 2d PSFs, and was used to quantify
the effects of such noise on PSF parameters across the GLAO field for
various seeing and noise conditions and zenith angles.  The
capabilities of the UD Monte Carlo code are summarized as follows:
\begin{itemize}
\item The atmospheric model can
cope with a large (not specifically-limited) number of
independently moving turbulent layers.
\item Multiple
laser beacons and/or Natural Guide Stars can be modeled.
\item Multiple DMs of a
number of types can be modeled.
\item Multiple Wavefront Sensors
(one per laser beacon or NGS) can be included. These include all
main detector noise effects, as well as the effects of detector
pixellation and atmospherically-induced speckle.
\item  The
science PSF may be sampled at a number of field points.
\end{itemize}

\subsection{Direct Comparison of Results}

Starting with the same inputs (Table 4), three figures
of merit were computed for PSFs compensated with GLAO: FWHM, 
ensquared energy within 0.1$^{\prime\prime}$, and Strehl ratio. 
Three turbulence
profiles were run both with and without photon and read noise
included.  In all models, spatial fitting error was included, and some
included WFS aliasing error. Other sources of residual wavefront error,
such as servo lag, were omitted. The intention was not to produce realistic
estimates of performance at {\it Gemini-S},
but merely to verify that the codes all
predicted essentially the same results (this was not true initially,
but excellent agreement was eventually achieved).

\begin{deluxetable*}{llllllll}
\tabletypesize{\small}
\tablewidth{0pt}
\tablecaption{Parameters of the common models used to validate the 
simulation codes.}
\tablehead{
Atmosphere & \multicolumn{7}{l}{$r_0=17$ cm at $\lambda = 500$nm. $L_0 = 30$ m} 
\\
 & \multicolumn{7}{l}{$C_n^2$ profiles (fractional power at each height):} \\
 & Height (m) & 0 & 300 & 500 & 900 & 2000 & 10000 \\
 & Profile 1: & 0.45 & 0.15 & 0.00 & 0.00 & 0.07 & 0.33 \\
 & Profile 2: & 0.30 & 0.00 & 0.30 & 0.00 & 0.07 & 0.33 \\
 & Profile 3: & 0.20 & 0.00 & 0.00 & 0.40 & 0.07 & 0.33 
}
\startdata
Telescope & \multicolumn{7}{l}{Outer Diameter = 8.0 m. No central obscuration} \\
\hline
FOV & \multicolumn{7}{l}{10$^\prime$ square} \\
\hline
Guide Stars & \multicolumn{7}{l}{5 NGS on a regular pentagon in a 
circle of radius $7^\prime.07$} \\
 & Brightness: & \multicolumn{6}{l}{Case 1: infinite} \\
 & & \multicolumn{6}{l}{Case 2: R=13 (85,400 photon/m$^2$/s)} \\
\hline
WFS & \multicolumn{7}{l}{10$\times$10 Shack-Hartmann, 8$\times$8 pixels 
per subaperture} \\
 & \multicolumn{7}{l}{Plate scale: 0.2$^{\prime\prime}$ per pixel} \\
 & \multicolumn{7}{l}{Read noise: 3.5 e$^-$ rms per readout per pixel} \\
 & \multicolumn{7}{l}{Frame rate: 500 per second} \\
 & \multicolumn{7}{l}{Wavelength: 700 nm monochromatic} \\
 & \multicolumn{7}{l}{No sky background} \\
\hline
DM & \multicolumn{7}{l}{Conjugate height: 0 m} \\
 & \multicolumn{7}{l}{Compensation: $\leq$ 77 degrees of freedom} \\
\hline
Test stars & \multicolumn{7}{l}{Co-ordinates: (0$^\prime$.0, 0$^\prime$.0) 
(2$^\prime$.5,
0$^\prime$.0) (5$^\prime$.0, 0$^\prime$.0) (2$^\prime$.5, 2$^\prime$.5)
(5$^\prime$.0, 5$^\prime$.0)} \\
 & \multicolumn{7}{l}{Wavelength: 1.25 $\mu$m monochromatic} \\
 & \multicolumn{7}{l}{Plate scale: 0$^{\prime\prime}$.1 per pixel} \\
 & \multicolumn{7}{l}{Integration time: $\geq$ 100 s} 
\enddata
\end{deluxetable*}

The results from the five codes are summarized in Figure 2 and demonstrate
agreement to within $\sim0.02^{\prime\prime}$ in the
PSF FWHM. The two Monte Carlo codes generally predict somewhat
worse performance than the three analytic codes, presumably
because of the inclusion in the Monte Carlo simulations of a greater 
range of physical effects.

\begin{figure}
\vbox to 3.3in{\rule{0pt}{3.3in}}
\includegraphics{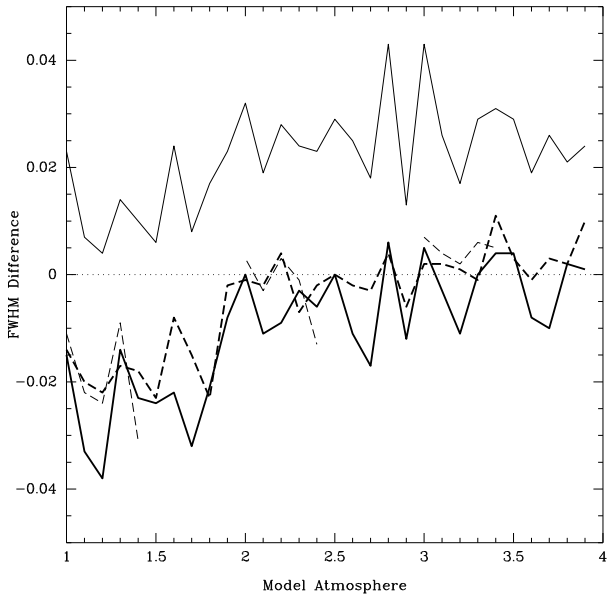}
\caption{Difference in simulated FWHM between various simulations
codes and the Durham Monte Carlo model for various model atmospheres, 
field positions,
and treatments of noise presented in Table 4. 
The UA Monte Carlo code results are marked with a solid line
(the residuals are all greater than 0), the UA Analytic code results 
are marked with a thin dashed line, the
PAOLA results are marked with a heavy dashed line, and the CIBOLA results
are marked with a heavy sold line. Results between 1-1.9 use Profile 1 from 
Table 4, results between 2-2.9 use Profile 2, and results between 3-3.9
use Profile 3. The ten PSF residuals shown for each profile are made
from five field positions with and without the inclusion of noise. 
Results labeled with numbers greater than or equal to one half 
for each profile came from simulations including noise. For an absolute
sense of scale, the Durham model FWHM for 1.0 is 0.235$^{\prime\prime}$, 2.0 
is 0.258$^{\prime\prime}$, and 3.0 is 0.307$^{\prime\prime}$. {\bf Regardless
of the details of the fits, model atmospheres, or inclusion of noise, the
five GLAO simulation codes produced FWHM which agreed to within 
0.04$^{\prime\prime}$. In most cases the agreement was even better.} The 
analytic codes produced virtually indistinguishable results; the Monte Carlo
codes, which included a greater range of physical effects, produced 
corrected FWHM which were not quite as narrow in most cases.}
\end{figure}

\section{Tools for Evaluating GLAO Performance}

The PSF from a GLAO system is very different from the PSF of a
diffraction-limited AO system. Therefore, the proper merit functions
for evaluating a GLAO system need to be identified. We begin this
process by describing the functional form of the GLAO PSF and
then present and discuss various performance metrics 
we use to gauge the performance of GLAO.

\subsection{The GLAO Point Spread Function}

The PSF produced by a wide-field GLAO system exhibits
no diffraction-limited peak, and qualitatively
is very similar to the PSF generated without any form of AO.
The shape of the GLAO PSF shown in Figure 3 is 
well fit by the same function commonly used to describe seeing limited
PSFs, the Moffat
function (Moffat 1969):
\begin{equation}
I(r) = I_0 [1+(r/\alpha)^2]^{-\beta}.
\end{equation}
A Gaussian profile matches the profile shape only to roughly the 50\%
of the peak flux, while the Moffat function provides a good fit
below 1\% of the peak height. Based on our simulated PSFs, both
the GLAO PSF and seeing limited PSF are well fit by a
Moffat PSF with $\beta$ between 2.5 and 4.5.

\begin{figure}
\vbox to 3.3in{\rule{0pt}{3.3in}}
\includegraphics{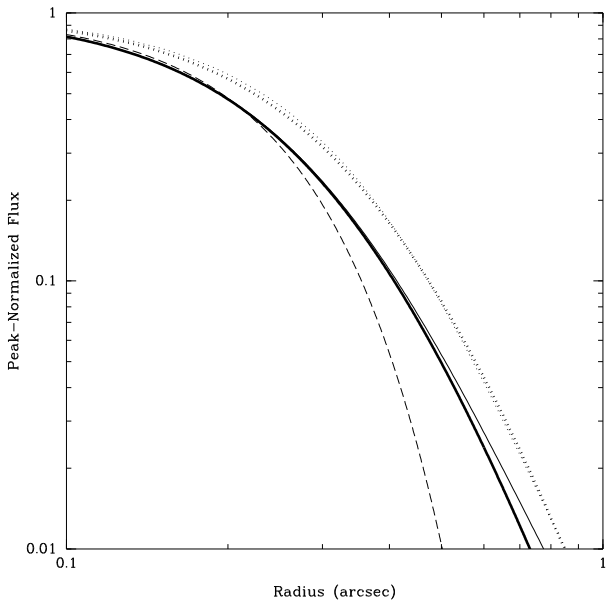}
\caption{Radial profile of GLAO corrected PSF in normalized flux versus radius
(Heavy solid line). The seeing-limited PSF is marked with a heavy dotted 
line (to
the right of the heavy solid line).  Gaussian (dashed line) and 
Moffat (thin solid line to GLAO PSF; thin dotted line to seeing-limited PSF) 
fits are marked as well. A Moffat function with $\beta=2.9$ provides 
an excellent fit
to both the GLAO and seeing-limited PSFs in this case.
The PSF was generated at a wavelength of 1.0 $\mu$m.}
\end{figure}
\subsection{GLAO Merit Functions}

As part of the Gemini feasibility study, a reasonably detailed
science case was prepared for GLAO. Science
requirements for GLAO with representative instruments were then defined. Proper
modeling of the science gains from a GLAO system require the full
PSF information. However, for many of the science cases, a few key
parameters were identified that gave a reasonable (but simplified)
understanding of the gains. We present and discuss several of these 
merit functions here:

\begin{description}
\item[Full-Width-Half-Maximum] (FWHM) of the PSF is a familiar and easily calculable
quantity.
For proper motion studies or work on crowded stellar
fields, at least with the expected GLAO PSF shapes, FWHM is a key
parameter.
\item[Half light radius] ($\theta_{50}$), or the radius enclosing 50\% of the
total energy,
is a particularly valuable
merit function if the shape of the PSF is not well understood.
A correlation exists between $\theta_{50}$ and FWHM (Figure 4), but because
of the variation in the Moffat parameter $\beta$, the half light radius may
be a better general merit function, since it is more tightly correlated with
the integration time ratio discussed below.
\item[Ensquared Energy] (EE) measured within an aperture (we used
0.1$^{\prime\prime}$ and 0.2$^{\prime\prime}$ apertures) is an important merit
function for spectroscopy as it indicates the amount of energy that enters a 
slit of a given size. The EE within 100 or 200 mas is small at most scientific
wavelengths, however, as shown
by Figure 5.
\item[Integration Time Ratio] (ITR) is defined as the ratio of 
required exposure times
to reach a given signal-to-noise ratio in the optimal aperture in the background
limited case without and with GLAO; the higher the value of ITR, the greater
the GLAO performance gain. ITR is especially relevant to science cases requiring
faint, point-like object imaging. As expected from the Signal-to-Noise
equation, ITR is proportional to the square of the 
ratio of seeing-limited to GLAO 
$\theta_{50}$ (Figure 6). 
\item[Image Quality Variation] over the scientific FOV
is an important scientific criterion
to consider. In practical terms, PSF uniformity across the FOV will make
observations easier to calibrate, reduce and interpret. As we will discuss
later, however, image quality variation is critically dependent on the
FOV being averaged (Figure 7). If areas near guide stars are included
in the measure of image quality variation, that variation will increase
substantially. In general, image quality is very uniform over interesting
FOVs.
\item[Strehl Ratio,] an important merit function for classical AO systems,
is defined as the ratio of the PSF peak flux to the peak flux
of the perfectly diffraction-limited PSF. However,
Strehl Ratio
has little meaning for a GLAO system; at the wavelength
of 1.25 $\mu$m, all codes predict less than 2\% Strehl ratios (Figure 5).
\end{description}

\begin{figure}
\vbox to 3.3in{\rule{0pt}{3.3in}}
\includegraphics{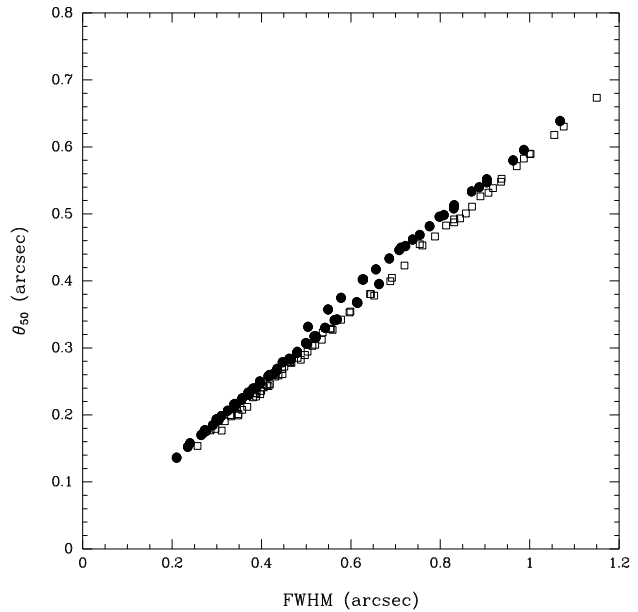}
\caption{FWHM versus radius enclosing 50\% of the total energy 
for a variety of wavelengths and turbulence profiles. The open squares are 
results for seeing-limited PSFs and the filled circles show results of 
GLAO-improved PSFs. The two parameters
are strongly correlated which indicates that the shape (i.e., the Moffat 
$\beta$ parameter) of the PSF does not
change significantly in these simulations. There are two sequences apparent 
in the GLAO PSFs. The points in the upper sequence were simulated using the
``Bad'' ground layer turbulence profile. In this case, it appears that the
core of the PSF is improved with GLAO, but that PSF halo is slightly stronger
which leads to a comparatively larger half light radius.
The overall tight relation between half light radius and FWHM means that
both parameters are valuable GLAO merit functions. 
}
\end{figure}

\begin{figure*}
\vbox to 6.5in{\rule{0pt}{6.5in}}
\includegraphics{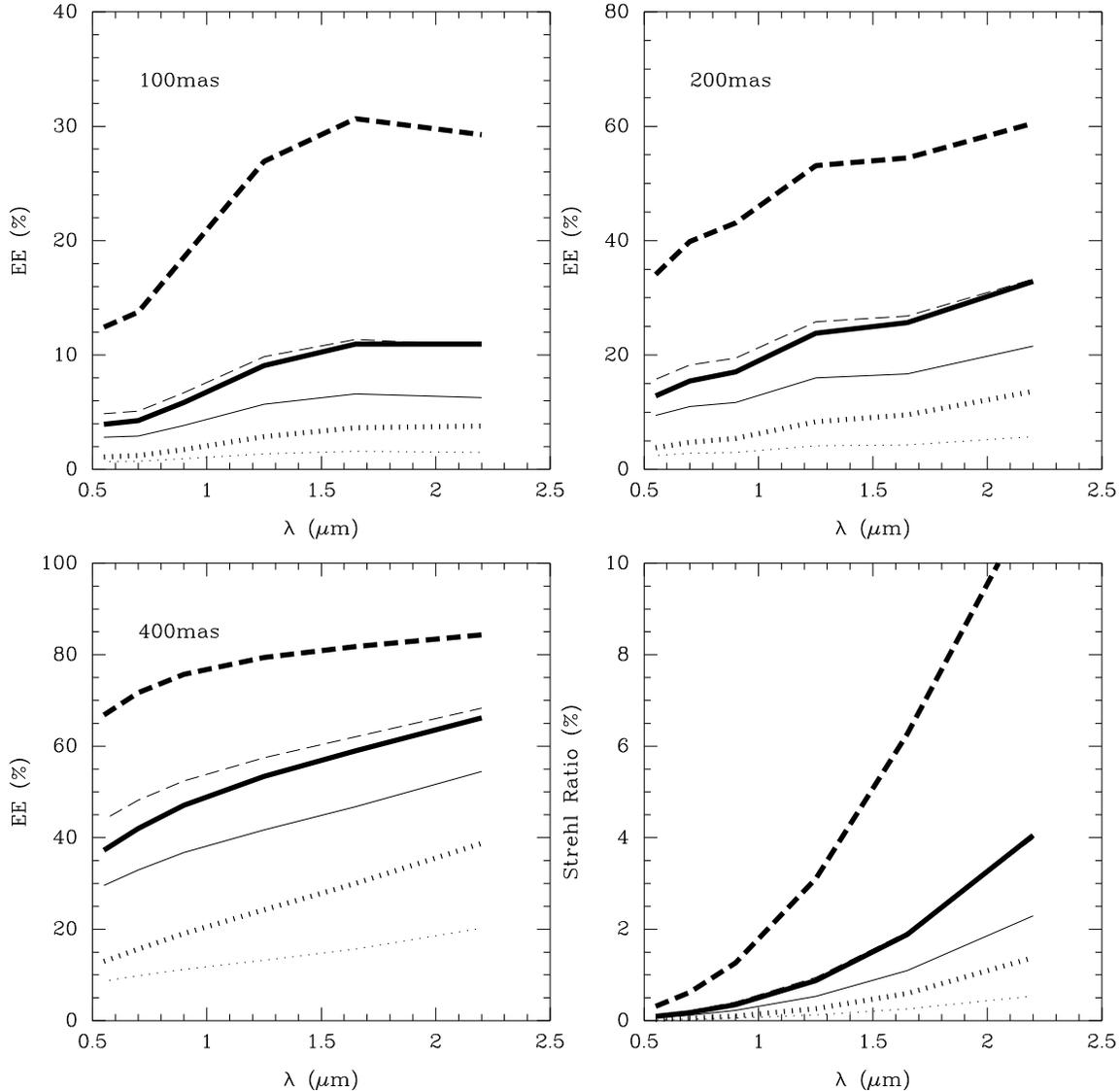}
\caption{Ensquared energy within 100, 200 and 400 milli-arcseconds and Strehl
ratio
versus wavelength for B:B (dotted lines), T:T 
(solid lines),
G:G (dashed lines). Heavy lines show the GLAO performance, while thin lines
show the seeing limited measurements. The ensquared energy within 100mas is 
less than 10\% at most wavelengths, and for most turbulence profiles. Only 
20\% of the ensquared energy is within 200mas in most cases. The Strehl Ratio
is very low for wide field GLAO observations; the Strehl ratio is less than
4\% in most cases (excepting the G:G performance at $\lambda>1.5\mu$m).}
\end{figure*}

\begin{figure}
\vbox to 3.3in{\rule{0pt}{3.3in}}
\includegraphics{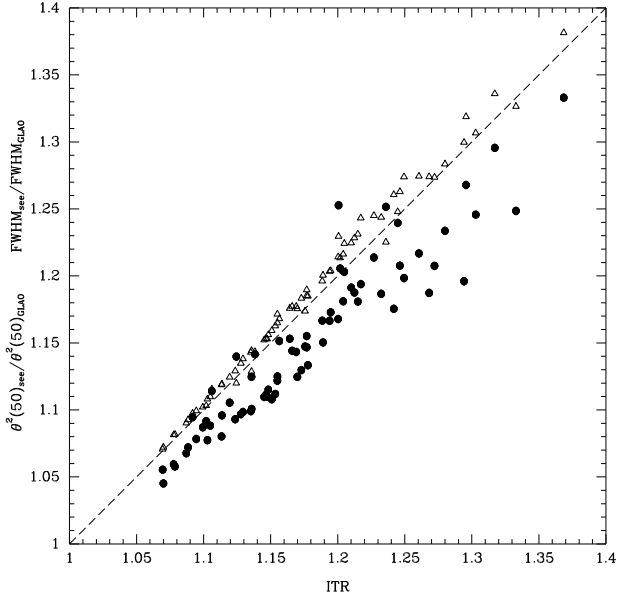}
\caption{Integration Time Ratio (ITR) versus the ratio of seeing-limited 
to GLAO-corrected FWHM (closed circles) and the square of the ratio of
seeing-limited to GLAO-corrected half-light radius $\theta_{50}$ (open triangles
)
for a range of wavelengths and turbulence profiles. While both FWHM and
$(\theta_{50})^2$ ratios are good proxies for ITR, it should be
noted that the correlation between
ITR and $(\theta_{50})^2$ ratio is significantly tighter.}
\end{figure}

\begin{figure*}
\vbox to 3.6in{\rule{0pt}{3.6in}}
\includegraphics{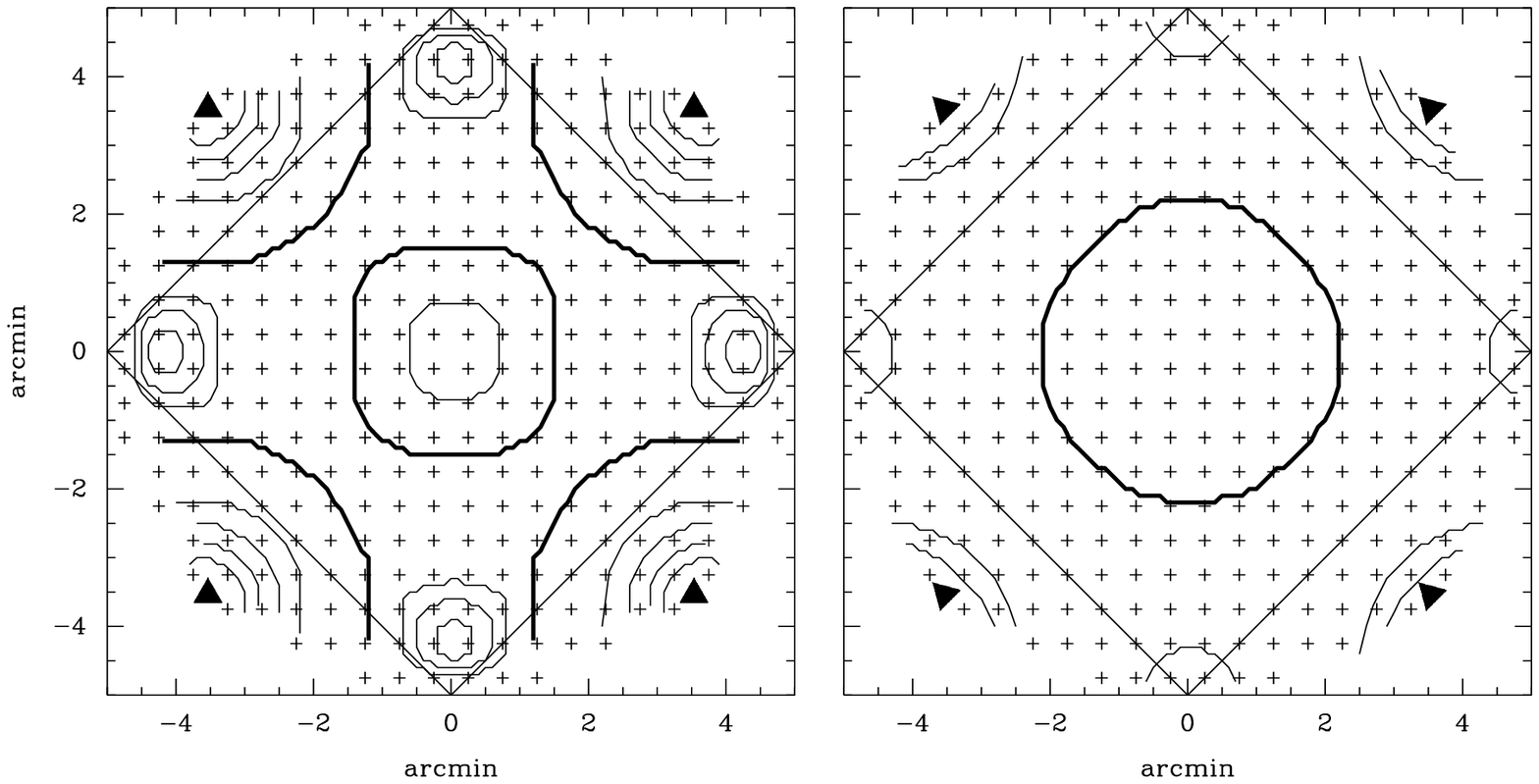}
\caption{Ensquared Energy in 0.2$^{\prime\prime}$ 
(EE; Left Panel) and FWHM (Right Panel) contours of 
GLAO PSF for the Good-Good turbulence profile as measured at 2.2$\mu$m.
A 7$\times$7 square arcminute FOV is marked,
as are each of the points simulated using PAOLA (+). LGS locations are 
marked with filled triangles. The performance is very uniform across
the FOV; the separations between contours is just 1\% in EE and 
0.01$^{\prime\prime}$ in FWHM. The thick lines
correspond to contours of lowest EE in 0.2$^{\prime\prime}$ (49\%) and 
largest FWHM (0.19$^{\prime\prime}$). 
Near a LGS, the wavefront error will be reduced because the average of the
pupils will be more heavily weighted by the turbulence from the free 
atmosphere in the direction of the LGS, thus leading to an improvement 
in image quality.  Then, as one moves away from the direction of a LGS beacon,
the performance will drop due to anisoplanatism.  However, in the 
direction between two
LGS, the total wavefront error will again decrease (and image quality improve)
because now the wavefront error due to free atmosphere
turbulence in that direction will be measured by two LGS WFS, and will not
cancel. In the center of the FOV, the image quality will again improve
very slightly as all 4 LGS WFS will sense a fraction of the turbulence
from higher layers. In essence, the ``grey zone'' is field-dependent and
is slightly higher in the field center, so more turbulence is corrected.
If the scientific
FOV can be chosen such that the LGSs are out of it, the correction across the
FOV will be very uniform which 
makes data reduction and calibration
easier.
}
\end{figure*}

For making general comparisons between the GLAO performance and
seeing limited performance, the FWHM or $\theta_{50}$ are the most useful
parameters. Ensquared energy is the most interesting merit function for
a specific spectrograph slit size, but 
the gains measured from ensquared energies
are very sensitive to the
size of the slit aperture; very small absolute GLAO performance gains will
be found if the aperture is significantly smaller than the
FWHM, because very little light will make it through a narrow slit aperture
in either the GLAO or seeing-limited case. The largest absolute and relative
GLAO performance 
gain is found if the aperture
size approaches the GLAO FWHM (Figure 5) which follows since the radial profile
of the GLAO PSF is falling at the given aperture slit width while the
radial profile of the seeing-limited PSF is still relatively flat and near
the peak value.
We therefore adopt the FWHM as the primary GLAO performance metric
through the course of this work, and will cite EE or ITR only when relevant.

\section{Performance of a Wide-field GLAO System}

Based on initial results of the simulations, that showed that performance of
GLAO systems are relatively insensitive to the specific LGS asterism and
FOV (see section 6),
a baseline GLAO configuration was adopted which 
employs four Sodium LGS arranged in a square, with each beacon
5 arcminutes from the center.  When used, we adopt
an asterism of three equally
bright NGSs arranged in a triangle (Figure 8; NGSs used for correcting
tip-tilt are only used in the Monte Carlo simulations).
We used WFSs with a relatively small number of
subapertures; only 10 to 17 samples across
the diameter of the DMs or WFSs were used (between 77 to 227 total subapertures)
depending on the simulation,
because a high order correction is not necessary to
achieve a good GLAO correction (See section 6.2.1). We averaged the signal
from the four LGS WFSs so the uncorrelated signal will cancel on average,
leaving only the common signal from the ground layer. The 
deviations in wavefronts
caused by layers over 2km will be uncorrelated. 
We modeled a GLAO system which employed an adaptive secondary mirror (first
proposed by Beckers 1989)
conjugated to -97m capable of
correcting between 80 and 230 modes depending on the WFS architecture
(-97m is the conjugate altitude of the current non-adaptive
secondary mirrors of the $\it Gemini$ telescopes.  As section 6.2.2 shows, the 
results of  simulations are relatively insensitive to the conjugate altitude.).
The performance of the GLAO system was modeled at four scientific wavelengths:
0.7 $\mu$m, 1 $\mu$m, 1.65 $\mu$m, 2.2 $\mu$m corresponding roughly to 
$RJHK$-bands. 

\begin{figure}
\vbox to 3.3in{\rule{0pt}{3.3in}}
\includegraphics{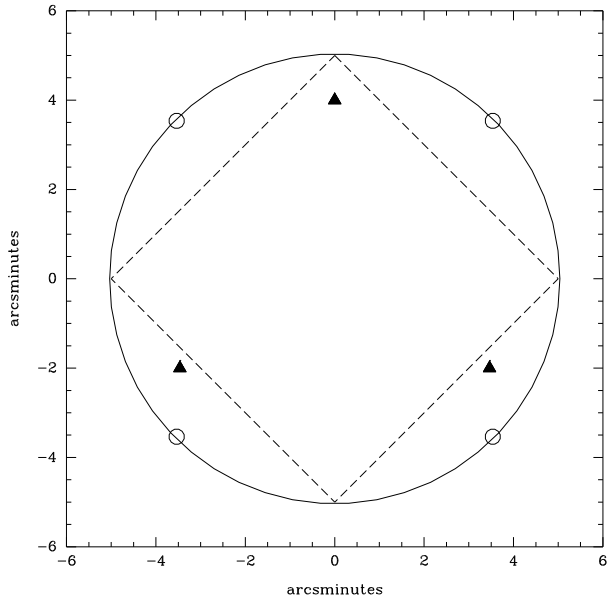}
\caption{The geometry of the baseline GLAO system consists of 4 LGS marked
with open circles on 5arcmin ring (solid line). 
3 NGS are marked with closed triangles, and a $7\times 7$ arcminute science 
FOV is marked with a dashed line.}
\end{figure}

\subsection{Image Quality Improvement}

At all wavelengths studied
and for most model atmospheres, we find that GLAO will decrease the
FWHM of a PSF by roughly 0.1$^{\prime\prime}$. However, 
the fractional change in PSF FWHM varies significantly from a factor of 3.8 
improvement in the
$K$-band with the Bad-Good profile, to just a factor of 1.1 improvement
in the $R$-band with 
the Typical-Bad profile.
The performance improvement
is greatest when the ground layer turbulence is large (Figure 9).
This means that the best image quality conditions 
which, without GLAO, occur only 20\% of the time occur 60-80\%
of the time with a GLAO system, transforming the cumulative distribution of 
image quality (Figure 10).
In particular, poor image quality
conditions occur only rarely once a GLAO system is employed. 

\begin{figure}
\vbox to 3.3in{\rule{0pt}{3.3in}}
\includegraphics{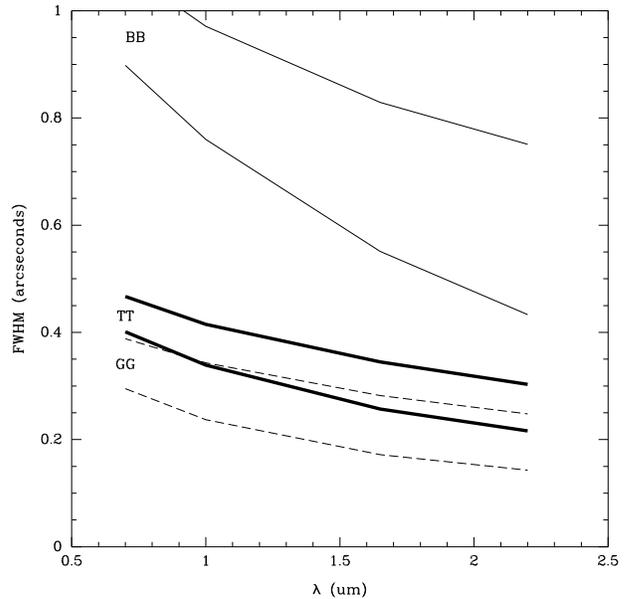}
\caption{FWHM versus wavelength, $\lambda$, for
three model atmospheres:
Good-Good (dashed lines), Typical-Typical (heavy solid line) and Bad-Bad
(thin solid line). The upper line is the seeing-limited FWHM and the lower
line is the GLAO-corrected FWHM. 
The correction is greatest for turbulence profiles with
Bad ground layers and at longer wavelengths. Simulations used a DM with 
77 degrees of freedom
and a 10 arcminute FOV.}
\end{figure}

\begin{figure*}
\vbox to 6.2in{\rule{0pt}{6.2in}}
\includegraphics{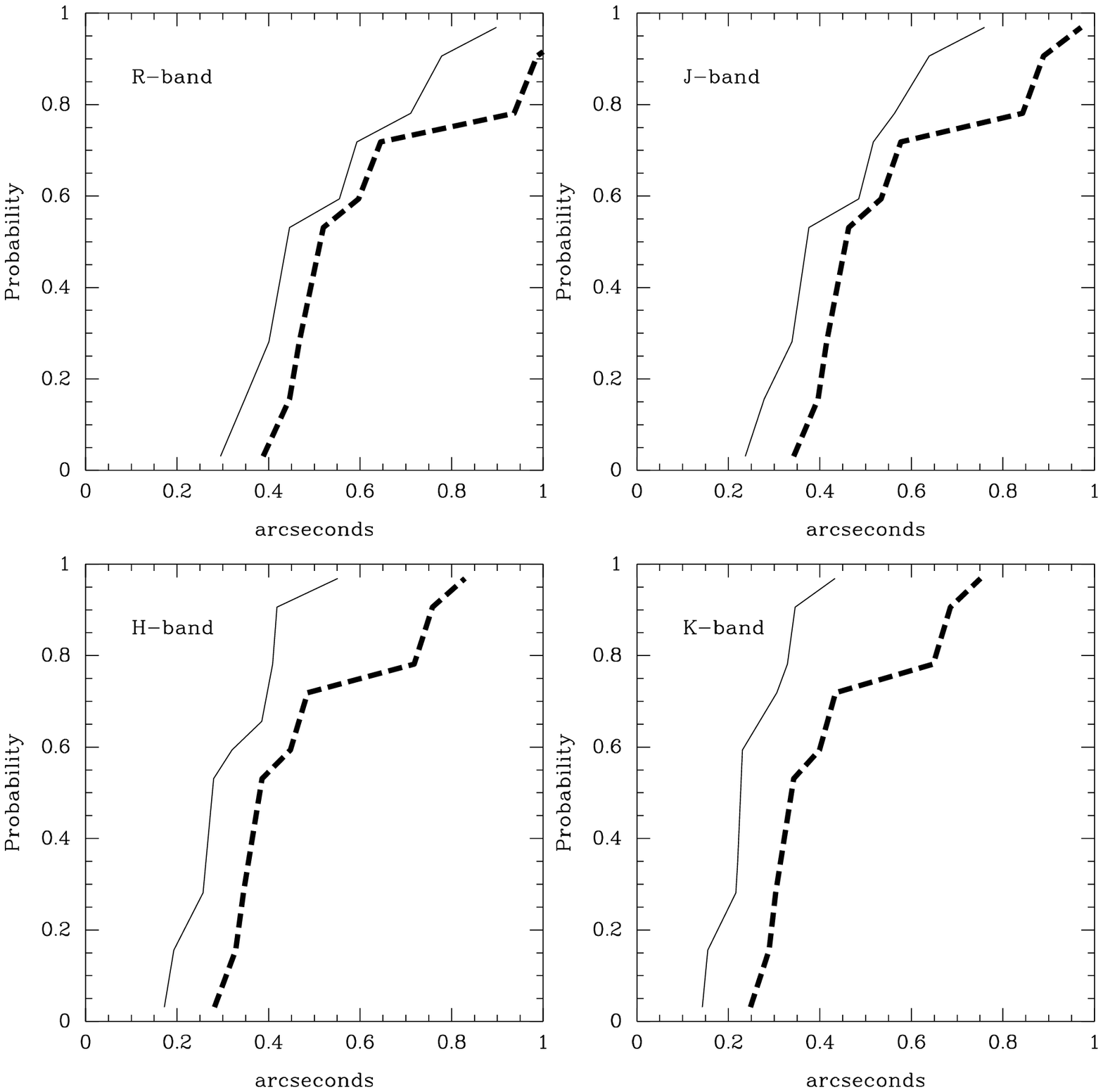}
\caption{Cumulative histogram of FWHM based on the nine model atmospheres
for both seeing-limited (heavy dashed lines) and GLAO (thin solid lines)
cases for wavelengths of $0.70\mu$m ($R$-band; upper left panel), 1.00$\mu$m
($J$-band; upper right panel), 1.65$\mu$m ($H$-band; lower left panel) and
2.2 $\mu$m ($K$-band; lower right panel). All simulations used a DM
with 77 degrees of freedom and a 10 arcminute FOV. A GLAO correction can alter
the image quality statistics at a site; the relatively greater improvement when
seeing is worst (and presumably the ground layer turbulence is greatest)
means GLAO can virtually eliminate bad-seeing nights; the poorest image quality 
occurring 30\% of the time without GLAO will only occur $\sim10$\% of the time
with GLAO.}
\end{figure*}

The expectation from previous studies (Rigaut 2002) was that
the $J$-band GLAO FWHM should be roughly 0.2$^{\prime\prime}$. As the
results in Figures 9 and 10 show, the results presented here are more 
pessimistic. Simulations of the PSF using only the
free atmosphere turbulence showed that even for a perfect GLAO correction
the FWHM is greater than 0.2$^{\prime\prime}$ under most atmospheric
conditions. We note that the full GLAO simulation of atmospheres with
``bad'' free atmospheres yield smaller FWHM than from simulations of
seeing-limited observations which included only the free atmosphere.
This is due to the relatively low altitude of the ``bad''
free atmosphere (3km), which is being partially corrected by the GLAO
simulations. Measured FWHM greater
than 0.4$^{\prime\prime}$ were initially attributed to a number of different
factors, but as we show in section 6, the results are relatively
insensitive to various trades. We believe the superior GLAO performance
quoted previously can be primarily attributed to the adoption of simpler, more
optimistic model atmospheres. The larger number of ground layers for our
model atmospheres combined with the probabilities of
given atmospheric conditions occurring, produce 
more realistic estimates of GLAO performance gains
(which are still significantly improved over the seeing-limited performance).

Despite these lower estimates of GLAO-corrected FWHMs provided by our models,
the gains in observing efficiency are still dramatic. Assuming background
limited imaging and an optimal point source extraction radius,
one can combine the GLAO gains using the model atmosphere
probabilities listed in Table 2 to estimate GLAO efficiency gains between
1.5 and 2.0 at
different scientific wavelengths (Figure 11). This can translate into
a substantial gain for an observatory; based on the Gemini Observatory
2004B proposal statistics, GLAO would benefit 55\% of the programs 
(proposals requesting observations between 0.6 and 2.2$\mu$m that
do not require high order AO) and would improve the efficiency
of the whole observatory by a factor up to 30\% to 40\%.

\begin{figure}
\vbox to 3.3in{\rule{0pt}{3.3in}}
\includegraphics{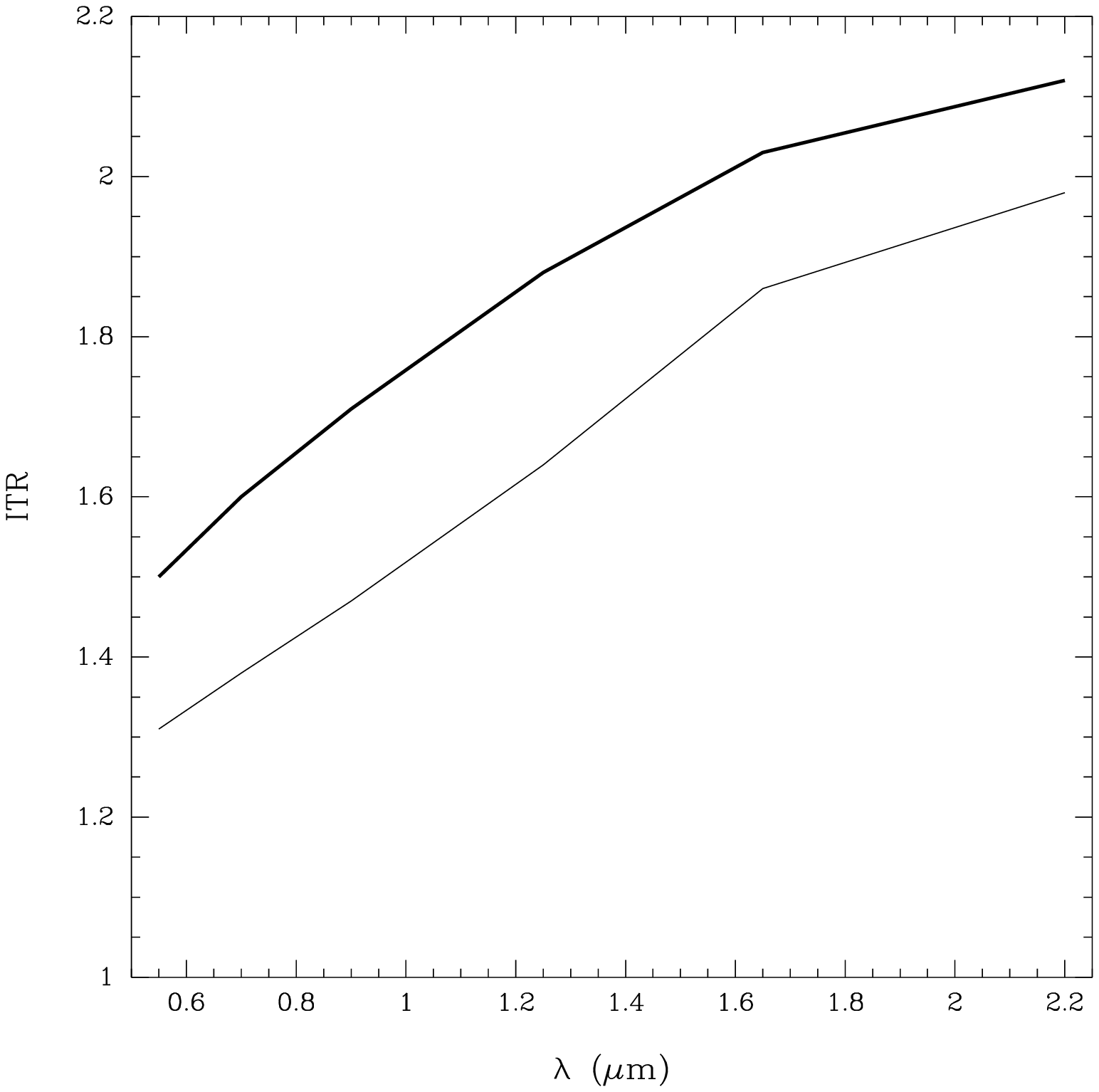}
\caption{Integration Time Ratio (ITR) as a function of wavelength using DMs
with 227 degrees of freedom (light line) and 2000 degrees of freedom 
(heavy line). This second, hypothetical DM with an enormous number of actuators
represents the limit of ITR gain with a GLAO system. ITR was calculated
by using a weighted sum over the nine turbulent profiles (See Table 2).
Increasing the
actuator density can significantly increase the performance of a GLAO
system.}
\end{figure}

\subsection{Performance Off Zenith}

The performance off-zenith was studied assuming bright natural guide
stars with the Typical-Typical atmospheric profiles.
Field quadrant averages and standard deviations
are given for all parameters measured at a wavelength of 1.6$\mu$m in Table 5.
As expected, the performance decreases off zenith. The GLAO-corrected
FWHM varies as a power-law of airmass with an exponent of 0.875, while
the power-law without AO is 0.6. GLAO performance will always degrade
with increasing airmass faster than the seeing-limited case because
GLAO includes fitting error. Fitting error increases at the same rate as 
seeing plus
anisoplanatism and will therefore degrade rapidly as more layers move into the
grey zone described by Tokovinin.

\begin{deluxetable}{lllll}
\tabletypesize{\small}
\tablewidth{0pt}
\tablecaption{Performance off zenith for Typical-Typical atmosphere for
a scientific wavelength of 1.6$\mu$m.}
\tablehead{
\multicolumn{1}{l}{Zenith} &
\multicolumn{2}{c}{FWHM} &
\multicolumn{2}{c}{0.2$^{\prime\prime}$ EE} \\
\multicolumn{1}{l}{Angle} &
\multicolumn{1}{l}{mean ($^{\prime\prime}$)} &
\multicolumn{1}{l}{rms ($^{\prime\prime}$)} &
\multicolumn{1}{l}{mean} &
\multicolumn{1}{l}{rms} 
}
\startdata
0$^\circ$ & 0.299 & 0.013 & 0.211 & 0.006  \\
30$^\circ$ & 0.338 & 0.012 & 0.177 & 0.006 \\
45$^\circ$ & 0.401 & 0.012 & 0.137 & 0.004 \\
60$^\circ$ & 0.548 & 0.015 & 0.083 & 0.003 \\
\enddata
\end{deluxetable}

\subsection{Laser Power Requirements}

The Durham Monte Carlo modeling tool was used to estimate the
laser power requirements for a GLAO system.  
Figure 12 shows that a minimum flux of approximately 50 detected photons
per WFS sub-aperture per detector integration will be required to achieve
close to optimal  correction. We note that
the model considered the LGSs to be point sources and did not 
include spot elongation. Treating laser beacons
as point sources results in simulated WFS spots which are too small and
are thus accurately centroidable with fewer photons, so the predicted 
AO performance at low light levels is overly optimistic.
With this caveat, we found that for 227 total subapertures, 
a sodium laser beacon and typical
sodium layer column density, this translates to a LGS  launch power less than
1 Watt. Even though this is a lower limit on the required laser power,
sufficient WFS photon flux should be achieved
with a relatively low power 2 to 5 Watt Sodium LGS system.
The main reasons why the required power is low compared to
other Laser AO systems is that the WFS signal is averaged, and
the error
budget for reaching the expected GLAO correction is relaxed; a GLAO system
does not produce diffraction-limited images even
in the near-IR so the required laser power is minimal.

\begin{figure}
\vbox to 3.1in{\rule{0pt}{3.1in}}
\includegraphics{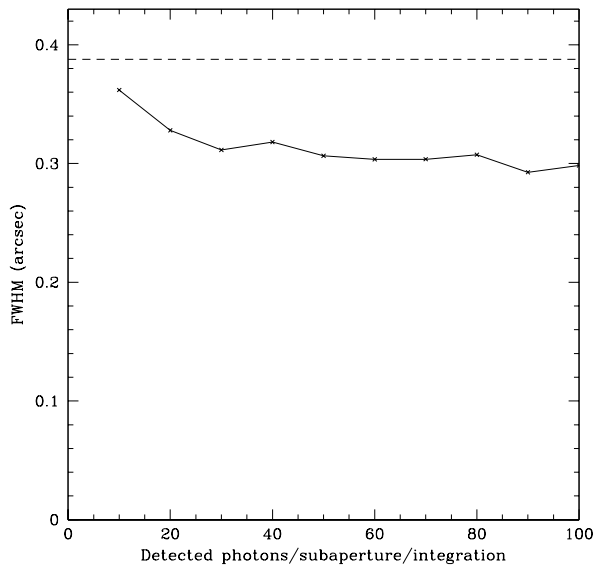}
\caption{FWHM of science PSF at 1.6 microns as a function of LGS photon flux.
Tip/tilt NGSs are assumed to be bright. The broken line shows the FWHM of the
uncorrected PSF. The Monte Carlo simulations assumed the LGSs were point 
sources.  Smaller spots are easier to centroid, therefore requiring
fewer photos per subaperture to be centroidable. Larger, elongated spots
will require more detected photons per subaperture, making this result
a lower limit on the required photos to be produced by the LGSs.}
\end{figure}

\subsection{Sky Coverage}

To compute the sky coverage expected for our baseline system we
used conservative estimates of the NGS noise performance drawn from
our Monte Carlo simulations (Figure 13), and assumed: 
1) 100 photons per integration
are required, 2) GLAO used a 500Hz sampling rate, 3) the overall
telescope plus detector efficiency was 60\% and 4) the WFS operated
in the V-band. Based on these assumptions, one needs 3 NGS stars with 
$V<15.0$\footnote{
The sampling rate of 500Hz was chosen only so that GLAO could be used to
remove potential telescope vibrations. If these vibrations are unimportant,
a sampling rate of 100Hz can be adopted which leads to a limiting 
magnitude for the NGS tip-tilt stars of $V<16.8$.}.
Based on the Bahcall and Soneira (1980) models of the Galaxy, there is
a density of $\sim$135 stars per square degree at the Galactic pole. For
each third of the 70 square arcminute FOV, there will be an average of
0.9 stars per sector patrolled by the NGS WFS. Assuming the number of stars are
Poisson distributed and spaced randomly over the FOV, there is a
20\% probability that all 3 NGS WFS probes can be placed on $V<15$ guide stars 
and a 81\% chance that at least one probe can be placed on a bright NGS and
the other two probes placed on fainter NGS read out at a slower rate (100 Hz).
Only one NGS needs to be bright enough
for the WFS to be read out rapidly
and control telescope
vibrations and wind shake;
the gain from having
3 NGS versus 1 NGS WFSs reading out at such a fast rate
will be significantly less substantial.

\begin{figure}
\vbox to 3.1in{\rule{0pt}{3.1in}}
\includegraphics{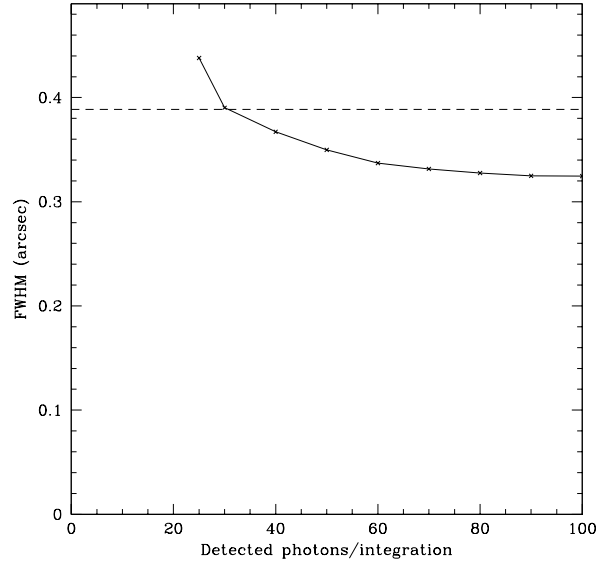}
\caption{FWHM of science PSF at 1.6 microns as a function of NGS (tip/tilt)
photon flux. LGSs are assumed to be bright. The broken line shows the FWHM
of the uncorrected PSF. A GLAO system requiring 100 photons/integration with
one WFS operating at 500 Hz translates into a 81\% sky coverage at the
North Galactic Pole (92\% sky coverage is achievable with slightly diminished
performance if only 60 detected photos per integration are required).}
\end{figure}

As shown in Figure 7, the non-uniformity of PSFs across the FOV is greatest
near the LGSs. NGSs have a lesser effect on the variation of PSFs (Figure 14);
the variation in PSFs introduced by NGSs is only apparent under the best
atmospheric conditions.
In general the PSF uniformity is still quite high over the selected FOV.

\begin{figure*}
\vbox to 3.6in{\rule{0pt}{3.6in}}
\includegraphics{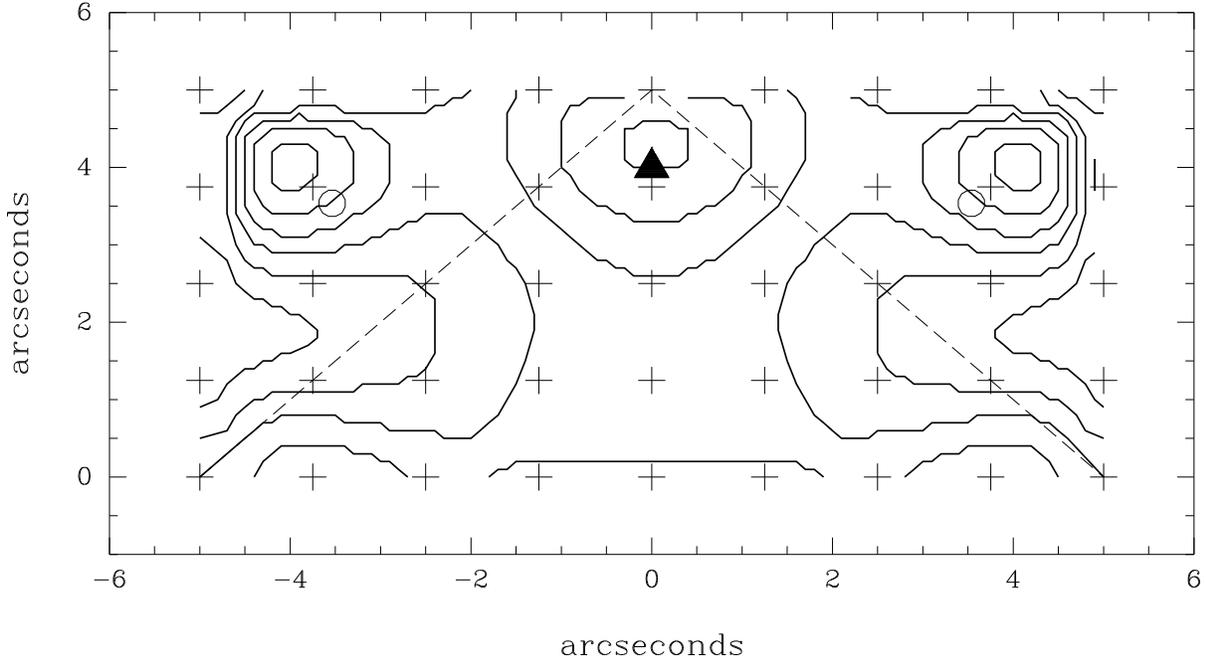}
\caption{Contour plot showing FWHM for Monte Carlo simulation incorporating
both LGS and low order NGS WFSs as measured at 1.65$\mu$m for the Good:Good
turbulence profile (which shows the strongest variations in FWHM at this
wavelength). LGSs
are marked by open circles, and the location of the NGS is marked with a
filled triangle. Locations of simulated points are marked with plus signs.
Contours are separated by 
0.05$^{\prime\prime}$ steps in FWHM. The GLAO correction is slightly
improved (by 0.1$^{\prime\prime}$) at the location of the
NGS when compared to the field center.}
\end{figure*}

\section{Trade Study Simulations}

Starting from this baseline model, we explore a large parameter space
and track GLAO performance.
We study how the performance of a GLAO system depends on the
corrected field of view, the DM actuator density (or equivalently in our
view, the WFS sampling), and different choices relating to the wavefront
sensors. Because we wanted to study these trades over a range of 
relevant scientific wavelengths and the nine model atmospheres,
we primarily used analytic modeling tools to carry out this work.

\subsection{Field of View Trade Study}

Our analytic simulations show that the GLAO performance does improve
as the FOV (i.e., the radius of the LGS asterism)
decreases, but the dependency between FOV and performance is weak;
the FWHM decreases by only 18\% when the area of the FOV is increased
by a factor of 6.25 (Figure 15). This is not too surprising as MCAO
systems, which also compensate for turbulence in discreet layers,
exhibit only weak dependencies on the size of the FOV (e.g. Le Louarn 2002).
The factor limiting GLAO performance gains is the strength of turbulence
in the free atmosphere; a simulation of the atmosphere excluding 
the ground layer (all layers under 1.6 km), showed that the mean seeing 
of a perfect GLAO system will only be 0.28 arcseconds for
a wavelength of 1$\mu$m assuming the 
simulated atmospheres and their weighted probabilities (Table 2)
are reasonable, compared to the
mean seeing without any adaptive optics of $0.56^{\prime\prime}$. 

\begin{figure}
\vbox to 3.3in{\rule{0pt}{3.3in}}
\includegraphics{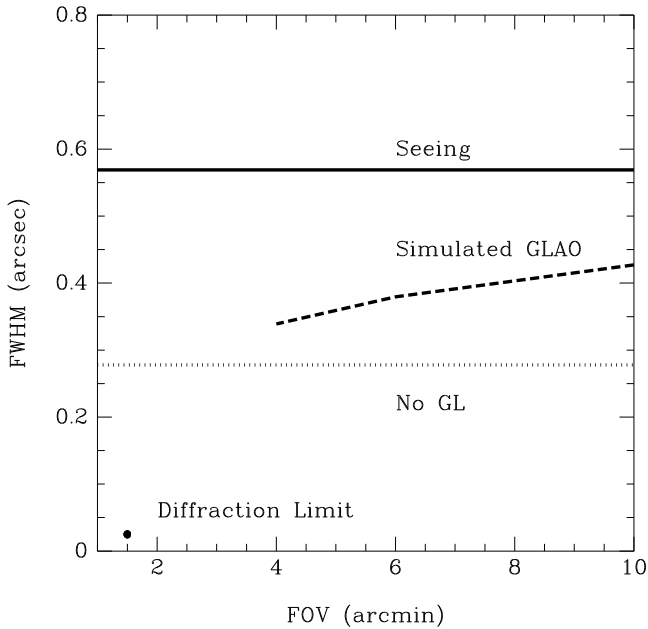}
\caption{FWHM as measured at 1$\mu$m as a function of the FOV using the 
Typical:Typical 
turbulence profile. While the GLAO performance improves as field size shrinks,
the gains are small; reducing the area of the FOV by a factor 
of 6.25 only 
improves the FWHM by 18\%.
}
\end{figure}

Even though the performance of a GLAO system decreases as the size
of the corrected FOV increases, the overall survey efficiency
(exposure time needed to survey a given area of the sky to a given
limiting magnitude) increases. Since many of the primary science cases for
MCAO and GLAO systems involve large surveys, we compared the relative survey
efficiencies for these AO systems. 
The proposed
Gemini GLAO imaging FOV is 49 square arcminutes compared to 2 square arcminutes
for GSAOI, the imager for the Gemini MCAO system. For this example,
the survey efficiency of a GLAO system is 4 times that of a MCAO system
for point sources. For non-point sources, the ratio of observing efficiency
increases dramatically; for objects with FWHM of 0.3 arcseconds, GLAO
has an observing efficiency 40 times that of MCAO\footnote{To be fair,
MCAO will yield better angular resolution for these objects enabling 
more science than a
mere detection.}. Because this measure
of survey efficiency
does not include acquisition and setup times, real gains in observing
efficiency are even greater when the additional overhead of
setting up 25 MCAO observations to cover the same FOV as a single GLAO
observation is taken into account.
For planned GLAO systems, this result
suggests that the GLAO FOV be made as large as possible, until the
extra acquisition overhead associated with running a LGS GLAO system
coupled to decreasing performance gain outweighs
the increased FOV.

\subsection{DM Property Trade Studies}

\subsubsection{Actuator Density}

Another important dimension of the GLAO parameter space which we studied
was the effect of varying actuator densities on GLAO performance.
Initial simulations used a DM with 77 degrees of freedom. If the actuator
density of the DM were significantly increased, we postulated that the 
performance may improve substantially because the fitting error would
decrease. 
We found that the optimal number of actuators depends on both the turbulence profile
and the scientific wavelength. In most cases the optimal number of
actuators is actually quite large ($\sim$30 actuators across the DM or 
$\sim 700$ degrees of freedom).
However, the performance in general is relatively
insensitive to the number of actuators (Figure 16). 
Only 314 degrees of freedom are needed to recover 95\% of the optimal
GLAO performance for $\lambda>0.7\mu$m. As figures 11 and 17
show, increasing the number of
actuators has the greatest relative impact at the shortest
scientific wavelength and when the free atmosphere has very little turbulence. 
In both these cases, fitting error dominates over other sources of error.
If the goal of a GLAO system were only to deliver improved performance in the
NIR, a DM with $\sim 80$ degrees of freedom would be adequate under 
most conditions.

\begin{figure}
\vbox to 3.5in{\rule{0pt}{3.5in}}
\includegraphics{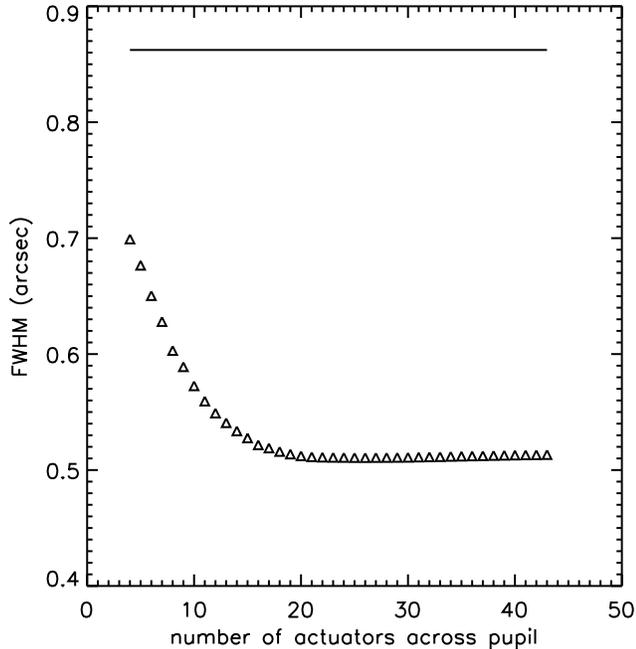}
\caption{As a function of the number of 
degrees of freedom across the diameter of a DM, 
the FWHM at the field center is plotted as triangles.
The solid horizontal line marks the seeing limited FWHM. Simulations were
performed using the Bad:Good profile and 4 LGS at a radius of 5 arcminutes at
a wavelength of 1 micron.}
\end{figure}

\begin{figure}
\vbox to 3.3in{\rule{0pt}{3.3in}}
\includegraphics{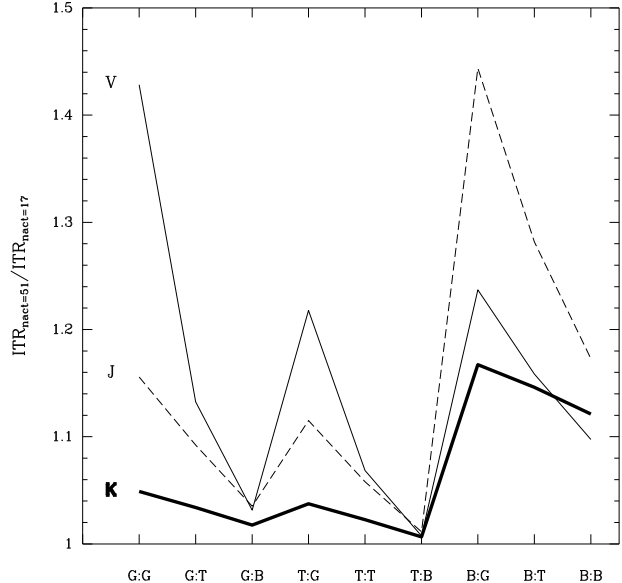}
\caption{Ratio of Integration Time Ratio (ITR) calculated using DMs with 
2000 (nact=51 across diameter) versus 227 (nact=17 across diameter)
degrees of freedom
as a function of the different turbulent profiles (labeled Ground
Layer:Free Atmosphere) for $V J K$-bands (0.55, 1.25, 2.2 $\mu$m). 
The gain at $K$-band is slight because 227 actuators are sufficient
to correct the ground layer turbulence for all profiles. This is not the case
in $J$-band, where it is clear 227 actuators are not sufficient to correct all
the ``Bad'' ground layer turbulence. Increased
actuator density improves GLAO performance for ``Good'' ground turbulence the
most at shorter wavelengths. Even 2000 actuators across the DM are probably 
insufficient to correct the ``Bad'' ground layer turbulence in $V$-band,
thereby limiting the gain in this regime.}
\end{figure}

\subsubsection{Conjugate altitude of DM}

If the DM in a GLAO system is conjugated to an altitude different than the 
effective height of the ground layer turbulence, one would expect 
anisoplanatism to degrade performance. However,
analytic simulations of a DM that is not precisely conjugated to the
ground layer atmospheric altitude show that the performance
does not suffer significantly; at worst a 5\% increase in FWHM is
observed for the Gemini telescopes (Figure 18). We find that for a configuration
of guide stars arranged in a pentagon, the optimal conjugation height 
is $\sim$100m considering
all results at 1 to 2.2$\mu$m and all nine model profiles. Comparing
the NGS pentagon to the LGS pentagon, we see that the constraint
on GLAO performance from DM conjugate height misregistration is
relaxed because of the
cone effect. This is an important result, because it means that 
adaptive secondary mirrors can be used with 
Cassegrain telescopes\footnote{The secondary mirror of a Cassegrain 
telescope is 
conjugated to below the primary mirror. For Gemini, the secondary is
conjugated to 97m below the primary.}, such as the 
{\it Gemini} telescopes,
and still produce 
GLAO performance gains.

\begin{figure}
\vbox to 3.5in{\rule{0pt}{3.5in}}
\includegraphics{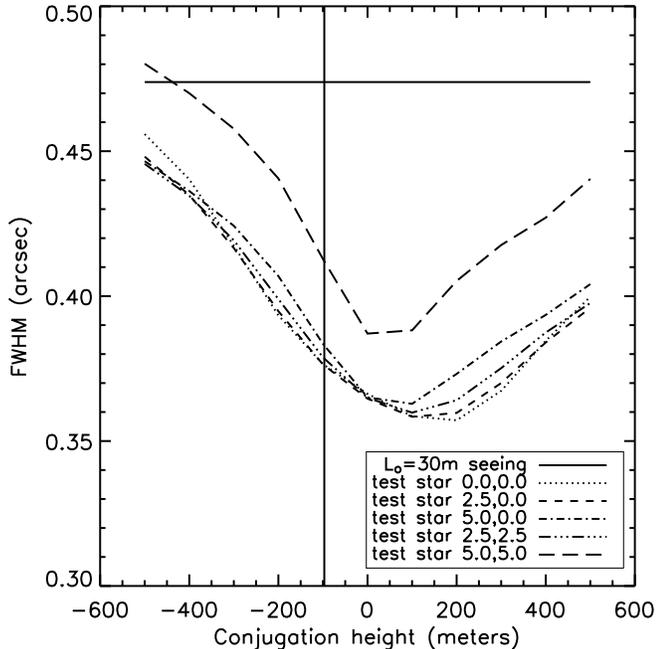}
\caption{As a function of DM conjugation height, the FWHM of 5 field positions 
plus the seeing limited case (see legend). FWHM were
measured from the Typical:Typical profile at a wavelength of 1 micron. 
Considering the five
field positions, the optimal conjugate altitude, depending on the 
image quality criteria, is around +100m. The conjugation of the Gemini
telescope secondary mirror 
is indicated by the vertical line at -97m, which suffers only a 5\% 
degradation in FWHM relative to the optimal FWHM for this case.}
\end{figure}

\subsection{Guide Star Trade Studies}

\subsubsection{Laser versus Natural Guide Stars}

We find that the GLAO performance is not optimal if all wavefront sensing
is done using NGSs.
For four real asterisms of NGSs near the North Galactic Pole, the mean and
standard deviation 
of the PSF FWHM were calculated. Variations in the PSF are 5\% greater
over the FOV if NGS versus LGS asterisms are used.

For one GLAO simulation using 3 NGSs, we looked at the morphology
of the PSF in greater detail. A comparison of the FWHM for the NGS
and LGS system show again that the uniformity of the PSF FWHM is
much higher for the LGS system (Figure 19); 
the standard deviation in FWHM is 34 mas in this NGS
GLAO simulation compared to 8 mas for the LGS GLAO simulation.
Furthermore, the correction yields an improvement in FWHM 
of only 0.05$^{\prime\prime}$, 
roughly half the 
correction achieved using the LGS system. 
We measured 
ellipticity (one minus the axis ratio) of the 
isophotes corresponding to the radius of 
the FWHM.
The magnitude and variation in ellipticity for the LGS 
simulation was small; the mean ellipticity is only 0.02. 
 The change in shape of the PSF is roughly 2\%.
With NGSs alone, ellipticity becomes more 
significant; a mean ellipticity of 0.10 is observed (Figure 19).
Ellipticity is a second order deviation --- ``higher order''
deviations to the PSF shape appear to be negligible. Perhaps
some of these disadvantages associated with using NGSs could be
alleviated if even more NGSs are averaged and an optimal reconstruction
algorithm were used (Nicolle \etal 2006).

\begin{figure*}
\vbox to 6.8in{\rule{0pt}{6.8in}}
\includegraphics{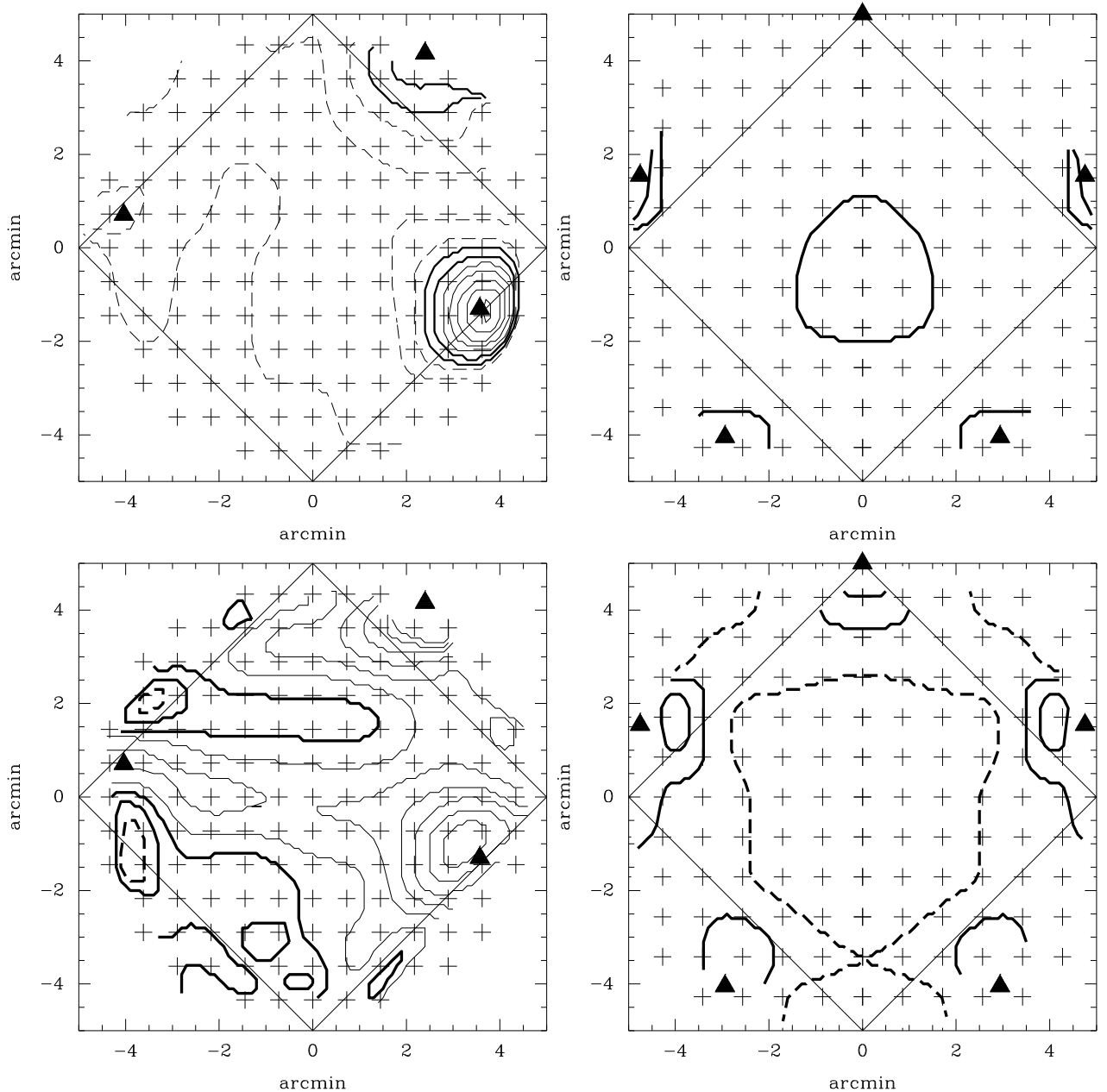}
\caption{Comparison of GLAO correction using three NGSs of different 
brightnesses scattered
randomly over the field (Left Panels) or five LGSs arranged in a
regular pentagon (Right Panels). For each guide star asterism (marked with
filled triangles), we show contours of FWHM (top panels) and 
ellipticity (bottom panels). The FWHM is smaller when LGS are used 
(both top panels have the same contours marked with heavy lines); the 
contours, separated by 0.02$^{\prime\prime}$, show the NGS correction is 
between 0.02$^{\prime\prime}$ and 0.04$^{\prime\prime}$ worse over the
majority of the FOV.
More importantly for PSF calibration, the shape of the PSF is
uniform when a regular asterism is used. In the ellipticity figures,
the lowest measured ellipticity contour ($1-b/a=0.02$ of the PSF measured at the
FWHM) is marked with a dashed solid line and is a good description for most
of the FOV when LGS are used. The ellipticity in the PSF is as high as 0.14
within the FOV for the NGS case (contours are separated by 0.02 in 
ellipticity). Locations of the simulated PSFs are marked by plus signs, and
a 7$\times$7 square arcminutes FOV is marked on each figure as well.
}
\end{figure*}

Of course one other major advantages of using LGS versus NGS WFSs is
that, as discussed in section 5.4, almost complete sky coverage
can be achieved for a GLAO system using NGS to correct only the
tip-tilt.

\subsubsection{Number and Geometry of Laser Guide Stars}

We explored GLAO performance for
a range of guide star numbers and
geometries. Cone effect and the altitude of the
beacons were not included,
which simplified the model and set aside the question of the relative
placement of the high-order beacons and tip-tilt beacons that must be
addressed with LGSs. Guided by the theoretical result of Tokovinin (2004)
that
the ideal beacon geometry for GLAO is a complete ring at the edge of the
FOV, regular polygons, with and without an additional axial beacon,
were explored from a triangle to a heptagon. For comparison, a single
axial beacon was also investigated.

As Table 6 shows,
more beacons yield slightly better results, as expected. Most of the
performance gains are obtained by going from 1 to 3 LGSs. The addition
of extra beacons only marginally improves the result when four or
more beacons are employed. This is consistent with results reported
from MCAO modeling (Fusco, \etal 1999). Adding more beacons in a GLAO
system means that turbulence from high layers cancels out better because
more non-overlapping high layer turbulence volumes are measured by the WFSs.
We suspect the cancellation of high layer turbulence in the mean wavefront
improves as the square root of the number of beacons.

\begin{deluxetable}{llll}
\tabletypesize{\small}
\tablewidth{0pt}
\tablecaption{Field-averaged FWHM in arcsec for each guide star
geometry using the Good:Good, Typical:Typical and Bad:Bad Cerro Pach\'on
turbulence profiles. Measurements were made for a
scientific wavelength of 1$\mu$m and a 10 arcminute diameter FOV. Guide
stars are evenly spaced around this diameter, unless the configuration is
denoted, e.g. 3+1, in which case one LGS is located at the center of the
FOV.}
\tablehead{
\multicolumn{1}{l}{Guide Star} &
\multicolumn{1}{l}{Good:Good} &
\multicolumn{1}{l}{Typical:Typical} &
\multicolumn{1}{l}{Bad:Bad} \\
\multicolumn{1}{l}{Geometry} &
\multicolumn{1}{l}{} &
\multicolumn{1}{l}{} &
\multicolumn{1}{l}{} 
}
\startdata
1       & 0.293$\pm0.027$ & 0.471$\pm0.043$ & 0.845$\pm0.070$ \\
3       & 0.230$\pm0.007$ & 0.382$\pm0.012$ & 0.727$\pm0.017$ \\
3+1     & 0.217$\pm0.008$ & 0.361$\pm0.014$ & 0.696$\pm0.025$ \\
4       & 0.221$\pm0.006$ & 0.368$\pm0.009$ & 0.708$\pm0.014$ \\
4+1     & 0.212$\pm0.007$ & 0.354$\pm0.011$ & 0.687$\pm0.021$ \\
5       & 0.215$\pm0.006$ & 0.359$\pm0.009$ & 0.694$\pm0.013$ \\
5+1     & 0.209$\pm0.006$ & 0.348$\pm0.010$ & 0.678$\pm0.018$ \\
6       & 0.212$\pm0.003$ & 0.354$\pm0.004$ & 0.688$\pm0.007$ \\
6+1     & 0.207$\pm0.004$ & 0.346$\pm0.007$ & 0.674$\pm0.015$ \\
No GLAO & 0.351 & 0.474 & 0.992 \\
\enddata
\end{deluxetable}

\subsubsection{Rayleigh versus Sodium Beacons}

A comparison of the GLAO performance was made where the only
difference was a change in beacon height appropriate for
Rayleigh and Sodium lasers. 
There is at best a 5\% improvement if Rayleigh
beacons are used, corresponding to a relatively constant $\sim$8 mas decrease
in FWHM for the model turbulence profiles. 
This advantage stems from the fact that the cone effect
from the lower Rayleigh beacon will be greater and thus be 
less affected by high 
turbulence layers. 

\subsubsection{Number of Tip-Tilt Stars versus Angular Resolution}

For a LGS WFS system, tip-tilt sensing is typically done with NGS.
Given the effects of anisoplanatism, a single star is inadequate to
correct the full GLAO field. A minimum of three is required to provide
compensation over the
field in both dimensions, but the question arises as to whether even more
will yield substantial improvements. Consequently, we investigated the level of
tilt correction with two guide star geometries: both used five Sodium LGS
on a circle of 10 arcmin diameter, with either 3 or 8 NGS arranged as a
regular polygon on the same circle. We found that 3 NGS are adequate
as the magnitude of the FWHM improvement by using 8 NGS was only 10\% 
greater than when 3 NGS were used (i.e. for a $\sim0.1^{\prime\prime}$ decrease
in the GLAO FWHM, increasing the number of NGS substantially decreased
the FWHM by just $\sim 0.01^{\prime\prime}$).

\section{Summary}

We have used the best available measures of ground layer turbulence profiles 
and a suite of modeling tools to study the performance of a GLAO system.
This work is the most complete study of GLAO to date, and our
results serve to ``de-mystify'' GLAO.
Among the many results of this study, we highlight the following:

$\bullet$ The shape of the GLAO PSF is qualitatively the same as a
seeing-limited PSF. Therefore, while having complete knowledge of the PSF
is desirable, the FWHM of the 
PSF is a practical and useful general metric for measuring GLAO performance.
Other performance metrics are more appropriate for specific applications, 
i.e., Integration Time Ratio is ideal for 
background-limited imaging and ensquared energy is the most useful metric
for spectroscopic observations.

$\bullet$ A GLAO system would significantly improve the image 
quality statistics. Unlike traditional AO systems, some of the greatest
gains to be had with GLAO are obtained when the seeing is poor.
In effect, this means that the number of nights with image quality
worse than the current 70\% level should be drastically reduced.

$\bullet$ Because diffraction limited imaging is not the goal of GLAO,
almost complete sky coverage is obtainable and the corresponding laser power 
requirements for a GLAO system are low.

$\bullet$ The performance of a GLAO system is relatively insensitive 
to a large number of trades. The performance is 
not a strong function
of the corrected FOV, the actuator density of the DM, the 
conjugate height of the DM, height of the LGS, or the guide star geometry.

$\bullet$ While our GLAO modeling of very wide fields showed that
performance gains are not as large as previously reported, the substantial
gains we do find would translate into major increases in the number of
scientific programs that can be completed in a given time on large telescopes.
Installing a GLAO system on a large telescope would increase the observing
efficiency by at most 40\%. 

$\bullet$ GLAO is highly complementary to other modes of AO. GLAO can improve
image quality in the optical and under intrinsically poor image quality
conditions where more traditional AO systems are unusable, especially if a
system with several hundred degrees of freedom is implemented. GLAO also yields
the greatest survey efficiency; as we have shown, the survey efficiency of
a GLAO system continues to increase with the FOV and should be much greater
than the MCAO survey efficiency. MCAO and classical AO could be used with
a GLAO survey instrument for follow-up observations of the most exciting
targets. In addition,
the sky coverage of a GLAO system will be greater than that of 
traditional AO systems due to the increased FOV and the insensitivity
of the performance to most of the studied variables.

\acknowledgements
The authors wish to thank S. Shectman for his useful comments and Gemini
Observatory and its staff for their contributions.

\bibliographystyle{apj}

\end{document}